\documentclass[aps,prl,twocolumn,letterpaper,superscriptaddress]{revtex4}
\usepackage{amssymb}
\usepackage{amsmath}
\usepackage{graphicx}
\usepackage{hyperref}

\def\E{\varepsilon}

\def\gd{G_{\mathrm{D}}}
\def\dgd{\delta G_{\mathrm{D}}}
\def\vd{V_{\mathrm{D}}}

\def\vc{V_{\mathrm{C}}}
\def\um{\mathrm{\mu m}}
\def\volts{\mathrm{V}}
\def\tesla{\mathrm{T}}
\def\mt{\mathrm{mT}}

\def\uv{\mu \mathrm{V}}
\def\uev{\mu \mathrm{eV}}
\def\dvd{\Delta\vd}

\def\Db{\Delta B}

\def\to{t_\mathrm{o}}
\def\ms{\mathrm{m}/\mathrm{s}}
\def\Vm{\mathrm{V}/\mathrm{m}}
\def\tp{\tau_\varphi}
\def\invtp{\tau_\varphi^{-1}}
\def\lphi{\ell_\varphi}
\def\invlp{\ell_\varphi^{-1}}
\def\perm2{\mathrm{m}^{-2}}

\def\etal{\emph{et al.}}
\def\FP{Fabry-P\'{e}rot }

\def\vdrift{v_{\mathrm{d}}}

\begin{document}
\title{Edge-State Velocity and Coherence in a Quantum Hall Fabry-P\'{e}rot Interferometer}
\author{D.\ T.\ McClure}
\affiliation{Department of Physics, Harvard University, Cambridge, Massachusetts
02138, USA}

\author{Yiming~Zhang}
\affiliation{Department of Physics, Harvard University, Cambridge, Massachusetts
02138, USA}

\author{B.\ Rosenow}
\affiliation{Department of Physics, Harvard University, Cambridge, Massachusetts
02138, USA}
\affiliation{
Max-Planck-Institute for Solid State Research, Heisenbergstr. 1, D-70569 Stuttgart, Germany}

\author{E.\ M.\ Levenson-Falk}
\affiliation{Department of Physics, Harvard University, Cambridge, Massachusetts
02138, USA}

\author{C.\ M.\ Marcus}
\affiliation{Department of Physics, Harvard University, Cambridge, Massachusetts
02138, USA}

\author{L.\ N.\ Pfeiffer}
\affiliation{Bell Laboratories, Alcatel-Lucent Technologies, Murray Hill, New
Jersey 07974, USA}

\author{K.\ W.\ West}
\affiliation{Bell Laboratories, Alcatel-Lucent Technologies, Murray Hill, New
Jersey 07974, USA}
\date{\today}

\begin{abstract}

We investigate nonlinear transport in electronic Fabry-P\'{e}rot interferometers
in the integer quantum Hall regime. For interferometers sufficiently large that Coulomb blockade effects are absent, a checkerboard-like pattern of conductance oscillations as a function of dc bias and perpendicular magnetic field is observed. Edge-state velocities extracted from the checkerboard data are compared to model calculations and found to be consistent with a crossover from skipping
orbits at low fields to $\vec{E}\times \vec{B}$ drift at high fields. Suppression
of visibility as a function of bias and magnetic field is accounted for by including energy- and field-dependent dephasing of edge electrons.
\end{abstract}

\maketitle

The electronic Fabry-P\'{e}rot interferometer (FPI), implemented as a quantum dot
in the quantum Hall (QH) regime, has attracted
theoretical~\cite{chamon97,bonderson06,stern06,bernd07,ilan08} and
experimental~\cite{caminoPRL05,camino06,caminoPRL07,godfrey07,willett08,yiming09}
interest recently, especially in light of the possibility of observing
fractional~\cite{chamon97} or
non-Abelian~\cite{stern06,bonderson06,ilan08,nayak08} statistics in this geometry.
Earlier experiments reveal that
Coulomb~\cite{vanwees89,alphenaar92,mceuen92} and Kondo~\cite{keller01,stopa03}
physics can play important roles, as well. With such a rich spectrum of physics in these devices, a thorough understanding of the mechanisms governing transport even in the integer QH regime remains elusive.

While most work on electronic FPI's to date has focused on transport at zero dc
bias, finite-bias measurements have proved to be a useful tool in understanding
the physical mechanisms important in other interferometer geometries. In
metallic~\cite{vanoudenaarden98} and semiconducting~\cite{vanderwiel03} rings
interrupted by tunnel barriers, oscillations in transmission as a function of magnetic field and dc bias, forming a checkerboard pattern, have been observed.
These features, attributed to the electrostatic Aharonov-Bohm (AB)
effect~\cite{aharonov59,nazarovPRB93,nazarovPB93}, were used to
measure the time of flight and dephasing in these devices. Similar
checkerboard-like lobe structures have also been observed in Mach-Zehnder
interferometers~\cite{neder06,roulleauPRB07,litvin08}. In that case, the pattern of oscillations is not readily explained within a single-particle picture and remains the
subject of continued theoretical
study~\cite{sukhorukov07,chalker07,levkivskyi08,nederPRL08}. In electronic FPI's,
conductance oscillations as a function of dc bias have been
investigated theoretically~\cite{chamon97} and provide a means of extracting the edge-state velocity from the period in dc bias. Edge-state velocity measurement without the use of high-bandwidth measurements~\cite{ashoori92,ernst97} will likely be useful in determining
appropriate device parameters to probe exotic statistics beyond the integer regime. This
approach was recently used~\cite{camino06} to measure the edge-state velocity at
$\nu=1/3$, though in a small ($\sim 1~\um^2$) device where Coulomb interactions, absent in the
theory, may be expected to play a dominant role~\cite{godfrey07,yiming09}.

In this Letter, we present measurements of finite-bias conductance oscillations in
an $18~\um^2$ electronic FPI whose zero-bias behavior is consistent with AB interference without significant Coulomb effects~\cite{yiming09}. We find a checkerboard-like pattern of conductance oscillations as a function of dc bias and magnetic field, in agreement
with the predictions of Chamon~\etal~\cite{chamon97}. Measuring the period in dc
bias allows the velocity of the tunneling edge state to be extracted over a range
of magnetic fields, yielding a low-field saturation consistent with a crossover
from $\vec{E}\times \vec{B}$ drift to skipping orbits. High-bias fading in the
checkerboard pattern is quantitatively consistent with a dephasing rate
proportional to energy and magnetic field. Zero-bias oscillations in a $2~\um^2$
device of similar design, where Coulomb effects are significant~\cite{yiming09}, do not evolve periodically with dc bias; instead, plots of conductance versus bias and magnetic field reveal diamond-like regions of blockaded transport in the weak-forward-tunneling regime that become more smeared out with stronger forward tunneling.

Devices are fabricated on GaAs/AlGaAs quantum-well structures with a
two-dimensional electron gas (2DEG) of density $n=2.7\times10^{15}~\perm2$ and
mobility $\mu = 2,000~\mathrm{m}^2/\mathrm{Vs}$ located $200~\mathrm{nm}$ below
the surface. Hall bars are wet-etched as shown in Fig.~1(a), and metal surface
gates are patterned by electron-beam lithography as in Fig.~1(b). Interferometers
are defined by negative voltages ($\sim -3\volts$) applied to all gates except
$\vc$, and samples are cooled in a dilution refrigerator to $\sim 20~\mathrm{mK}$.
A current bias $I$, consisting of a dc component of up to $30~\mathrm{nA}$ and a 135-Hz component of $400~\mathrm{pA}$, gives rise to the diagonal voltage $\vd$ across the device, measured directly across the width of the Hall bar [Fig.~1(a)]. Lock-in measurements of diagonal conductance, $\gd\equiv
dI/d\vd$, are used to study changes in interferometer transmission as a function
of both $\vd$ and perpendicular magnetic field $B$. As shown in Fig.~1(c), the
current-carrying chiral edge states can be partially reflected at each
constriction, leading to interference between the different possible trajectories
as a function of the phase accumulated by encircling the interferometer.

\begin{figure}[t]
\center \label{figure1}
\includegraphics[width=2.8in]{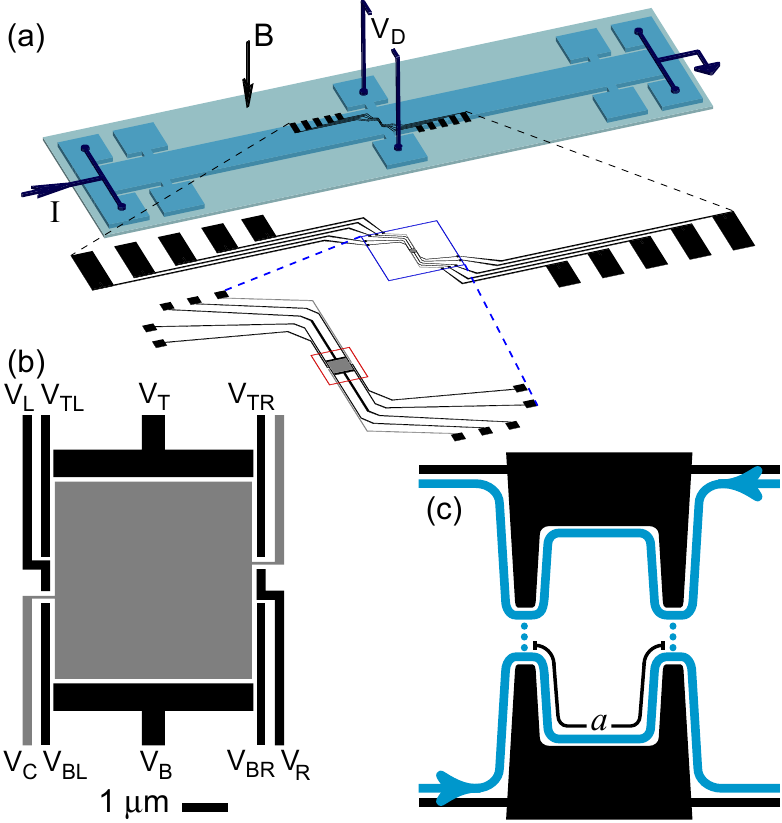}
\caption{\footnotesize{(a) With a current bias $I$ applied at one end of the Hall
bar, voltage $\vd$ is measured directly across its width. Surface gates are shown in increasing detail, with a red box indicating the region shown in (b). (b) Gate layout of the $18~\um^2$ device, which is operated as an interferometer by depleting all gates except $\vc$. (c) Schematic diagram of possible transmission paths through the device in the quantum Hall regime.}}
\end{figure}

A typical measurement of $\gd$ as a function of $B$ and $\vd$ in the $18~\um^2$ device is shown in Fig.~2(a), where a smooth background has been subtracted. A checkerboard-like pattern of oscillations
periodic in both $B$ and $\vd$ is observed, with reduced amplitude at high bias.
Similar patterns are seen at fields $B=0.22-1.26~\tesla$; over this range the
Landau level index, $N$, of the tunneling edge ranges from 4 to 1, but the field
period of oscillations is always $\Db \approx 0.25~\mt$, independent of both field
and bias.

\begin{figure}[h]
\center \label{figure2}
\includegraphics[width=2.6in]{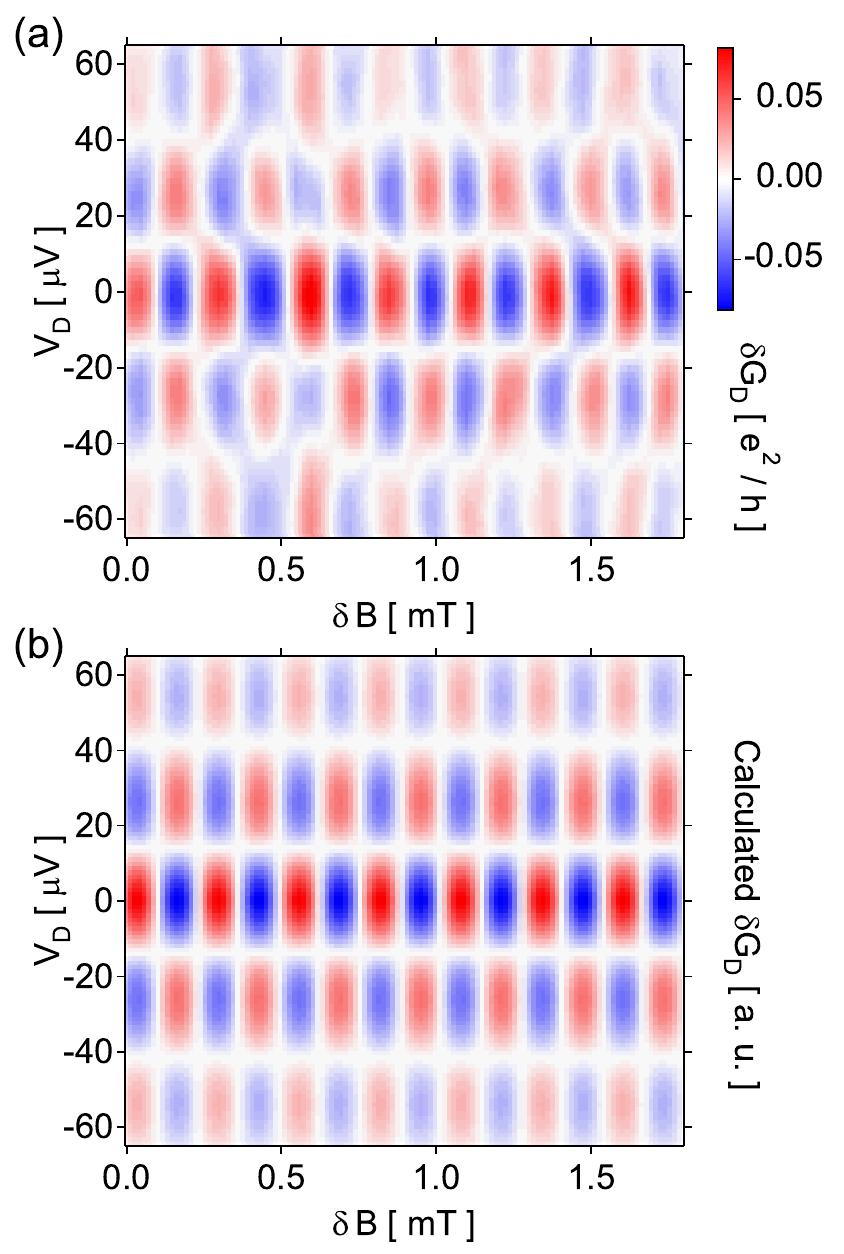}
\caption{\footnotesize{(a) $\gd$ as a function of $B$ and $\vd$ in the $18~\um^2$
device near $B=0.47~T$, with a smooth background subtracted. (b) $\dgd$ calculated
from Eq.\ (1), multiplied by the damping factor from Eq.\ (2), with $\Db=0.25~\mt$,
$\dvd=56~\uv$, and $\alpha=0.2$.}}
\end{figure}

Magnetoconductance oscillations in this device reflect AB interference of partially transmitted edge states \cite{yiming09}, with a phase shift $\Delta\varphi=2\pi\Phi/\Phi_0$, where $\Phi=BA$
is the flux enclosed (in area $A$) by the interfering edge, and $\Phi_0\equiv h/e$ is the magnetic flux quantum. The
observed field period corresponds to $A\approx 17~\um^2$, consistent with the dot area after subtracting a depletion length of roughly the 2DEG depth. The sinusoidal
lineshape of the oscillations seen here suggests that coherent transport is
dominated by two trajectories that differ in length by one traversal of the dot perimeter.

When a dc bias is added to $\vd$, an additional phase shift appears between interfering trajectories, associated with the energy-dependent wave vector of the contributing edge-state electrons; we will refer to this as the Fabry-Perot phase. The wave vector changes with energy as $\delta k = \delta \E / \hbar v$, where $v$ denotes the edge-state velocity. Following the analysis of non-interacting electrons in Ref.\ \cite{chamon97}, in which bias is assumed to affect mainly the chemical potential, we assign an additional relative phase $2 a  \E/\hbar v$ to an electron traversing the perimeter at energy $\E$ above the zero bias Fermi level, where $a\sim2\sqrt{A}=8.2~\um$ denotes the path length between constrictions [Fig.~1(c)]. For a symmetrically applied dc bias (relative to the gate voltages), and neglecting contributions from multiply-reflected trajectories, the expected differential conductance has the form
\begin{equation}
\dgd(\Phi,\vd)=\delta G_0\cos(2\pi\,\Phi/\Phi_0)\cos(eV_D a /v\hbar),
\label{checkerboard.eq}
\end{equation}
where the amplitude $\delta G_0$ does not depend on field or dc bias. Note that in this model, the contributions of  AB and \FP phase separate  into a product of two cosines, yielding a checkerboard pattern, as observed in the experimental data, Fig.~2(a). Ref.\ \cite{chamon97} predicts that when the bias is only applied to one contact, with the other contact held at ground (again, relative to the gates), the two phase contributions from bias and field instead appear as arguments of a single cosine, yielding a diagonal stripe pattern. Experimentally, the bias is always applied only at one end of the Hall bar, with the other end grounded; however, interaction effects within the dot are likely to effectively symmetrize the applied bias \cite{bert-priv}. Alternatively, a model in which the bias mainly affects the electrostatic (rather than chemical) potential~\cite{bernd-unp} also yields Eq.\ (1) without the need for a symmetric bias.  In either interpretation, the bias period corresponds to the edge velocity via $\dvd=(h/e)(v/a)$.

 We account for the reduced amplitude of oscillations at high bias by multiplying the right side of Eq.\ (1) by a damping factor, $e^{-2\pi\alpha\,|\vd|/\dvd}$, where $(2\pi\alpha)^{-1}$ gives the number of periods over which the amplitude falls to $1/e$ of its zero-bias value. Lacking theory for edge-state dephasing in FPI's, this form is motivated by the observation in related experiments of a dephasing rate proportional to
energy~\cite{vanderwiel03,roulleauPRL08}. We thus identify a voltage-dependent dephasing rate, $\invtp(\vd)=\alpha|e\vd|/2\hbar$, which reduces amplitude by
$e^{-2\to/\tp}$, where $2\to = 2 a /v $ is the time of flight around the
interferometer. To extract interference and dephasing parameters, the form
\begin{equation}
\delta G(\vd)=\delta G_0 e^{-2\pi\alpha\,|\delta x|}\cos(2\pi\,\delta x),
\end{equation}
where $\delta x=(\vd-V_\mathrm{off})/\dvd$ and $V_\mathrm{off}$ is a bias
offset, is fit to cuts of the data in Fig.~2(a), which yields a period $\dvd=56~\uv$ and dephasing parameter $\alpha=0.2$. These values, along with $\Db=0.25~\mt$ are then used
to produce the plot shown in Fig.~2(b). Figures~3(a) and 3(b) show
vertical cuts from data along with fits of Eq.\ (2) at $B=0.22~\tesla$ and
$1.26~\tesla$, respectively, representing a trend toward smaller $\dvd$ and larger
$\alpha$ at higher fields, the details of which we now study.

\begin{figure}[h]
\center \label{figure3}
\includegraphics[width=3.2in]{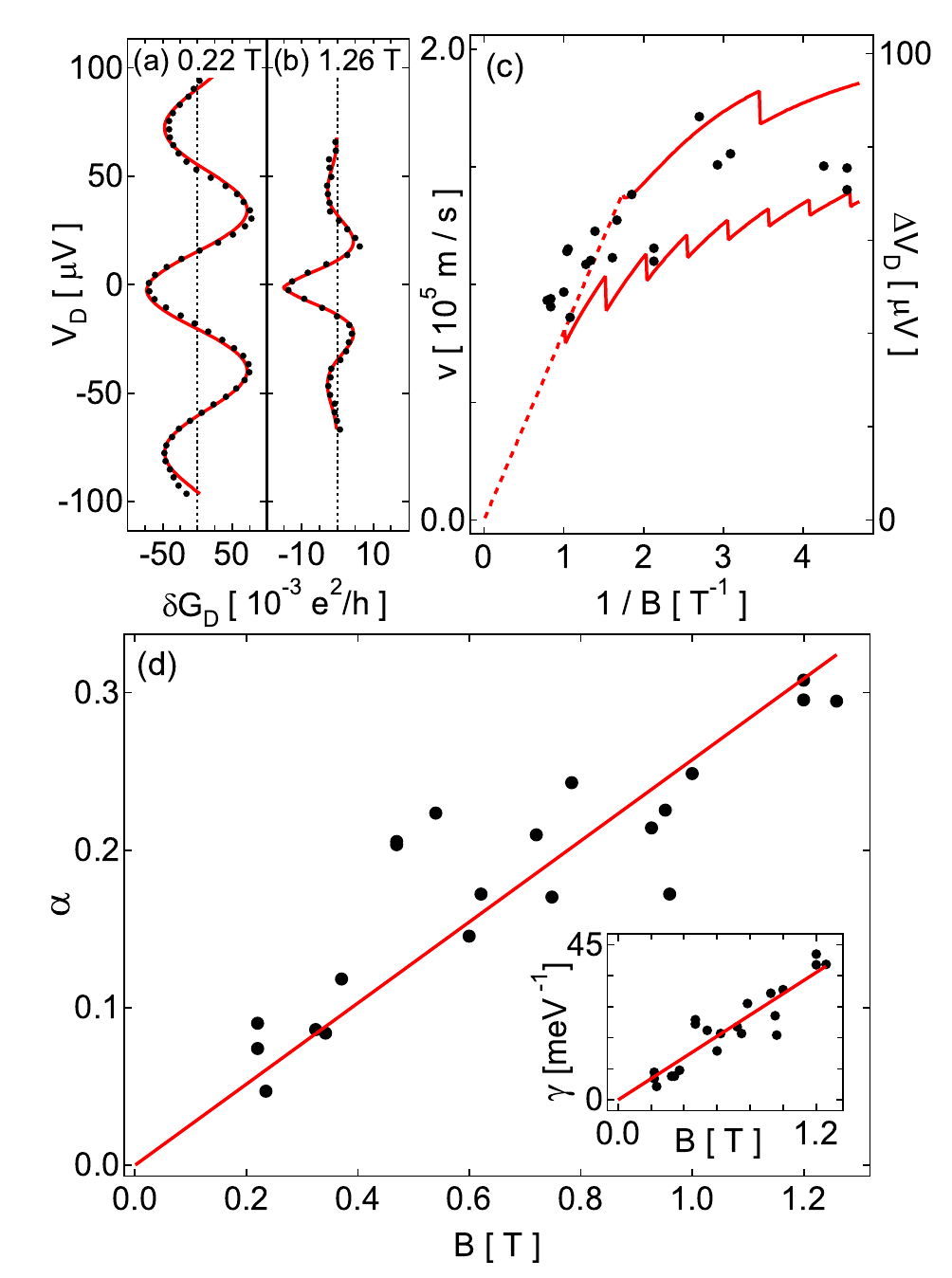}
\caption{\footnotesize{(a) $\gd$ as a function of $\vd$ (black dots) at a field of
$B=0.22~\mathrm{T}$, with a fit of Eq.\ (2) (red curve) yielding $\dvd=76~\uv$ and
$\alpha=0.063$. (b) Same as (a) but at $B=1.26~\mathrm{T}$ and yielding
$\dvd=47~\uv$ and $\alpha=0.34$. (c) Black dots indicate edge velocities (left
axis) determined from measured $\dvd$ (right axis) as a function of $1/B$. The red
curves indicate theoretical calculations: at low $1/B$, the diagonal dashed line
indicates the drift velocity corresponding to $E=8\times 10^4~\Vm$; at high $1/B$,
the top and bottom solid curves indicate the predicted skipping-orbit velocities
corresponding to the lowest and highest constriction densities, respectively. (d)
Best-fit damping parameter $\alpha$ as a function of $B$, with a linear fit of
slope $0.26~\mathrm{T}^{-1}$ constrained through the origin. Inset: $\gamma=
2\pi\alpha/e\dvd$ as a function of $B$, with a linear fit of slope
$31~(\mathrm{meV}\cdot\mathrm{T})^{-1}$ constrained through the origin.}}
\end{figure}

The black circles in Fig.~3(c) indicate the best-fit $\dvd$ (right axis) and
corresponding edge velocity (left axis) as a function of $1/B$. The velocities
appear roughly proportional to $1/B$ before saturating at $v\sim 1.5\times
10^5~\ms$ for $1/B \gtrsim 2~T^{-1}$. Red curves indicate calculations based on
single-particle models of edge velocities in two regimes. In the high-field limit,
where the cyclotron radius is much smaller than the length scale on which the
confining potential changes by the cyclotron gap, $\vec{E}\times \vec{B}$ drift
gives a velocity $\vdrift=E/B$, where $E$ is the local slope of the confining
potential. The data in this regime are consistent with a value $E\sim 8\times
10^4~\Vm$, which is reasonable given device parameters. At low fields, where the cyclotron radius exceeds the length scale set by $E$, electron velocities can be estimated from a skipping-orbit
model. For hard-wall confinement, the skipping velocity would be proportional to the cyclotron frequency and radius: $v_s\sim \omega_c r_c$.  Here, we have performed a detailed semi-classical calculation assuming a more realistic confining potential that vanishes in the bulk and grows linearly near the edge. In this regime, the predicted velocity depends on not only $B$ and $E$ but also on the Landau level index, $N$, resulting in a discrete jump in velocity for every change in $N$. Since the density in the constrictions (which along with $B$ determines $N$) varies over the course of the experiment, two theoretical curves are plotted in this regime: the
top one corresponds to the lowest observed constriction density of $2.8\times
10^{14}~\perm2$, and the bottom one corresponds to the highest, $9.5\times
10^{14}~\perm2$, both estimated from $\gd$ and $B$.

Figure~3(d) shows the best-fit damping parameter $\alpha$ as a function of $B$,
revealing rough proportionality: a straight line constrained to cross the origin
describes the data well with a best-fit slope of $0.26~\mathrm{T}^{-1}$. In
analogy to dephasing in 2D diffusive systems~\cite{altshuler85}, we
suggest that coupling to compressible regions in the bulk may lead to dephasing
with the $V_D$-dependence $\invtp\propto R_\square V_D$, where $R_\square$ is the
resistance per square in the bulk. Over the field range of our data, the bulk
longitudinal resistivity $R_{xx}$ (not shown) is on average roughly proportional
to $B$; taking $R_{xx}$ as an estimate of $R_\square$ would then lead to a
predicted dephasing rate proportional to both energy and magnetic field,
consistent with the data. Despite this agreement, we emphasize that Ref.~\cite{altshuler85} was not developed for edge states or FPI's, and a theory of dephasing in this regime remains lacking.

Alternatively, defining the damping factor as simply $e^{- \gamma |e V_D|}$, one
also finds rough proportionality between $\gamma$ and $B$, as shown in the inset
of Fig.~3(d). Here the best-fit slope for a straight line constrained through the
origin is $31~(\mathrm{meV}\cdot\mathrm{T})^{-1}$. The damping parameter $\gamma$ is related to $\alpha$ and to the dephasing length, $\lphi=v\tp$, by  $\gamma = \alpha\to/\hbar = 2 a/|e V_D| \ell_\varphi$; therefore, since $\to$ varies with field, at most one of $\alpha$ and $\gamma$ can be proportional to B. Physically, the latter case would correspond to $\invlp$ being the quantity that is linear in $B$ instead of $\invtp$. Experimental scatter prevents us from distinguishing these two possibilities.

Measurements on a $2~\um^2$ device of similar design, whose zero-bias
oscillations have previously been demonstrated as consistent with
Coulomb-dominated behavior~\cite{yiming09}, do not yield regular oscillations as a
function of bias. Figure~4 shows $\gd$ as a function of $B$ and $\vd$ in a regime
of weak forward-tunneling, where diamond-like features appear. Interpreting these
features as the result of Coulomb blockade yields a charging energy of roughly
$25~\uev$, reasonable given the device size, 2DEG depth, and the large capacitance
afforded by the top gate. In regimes of stronger forward tunneling, the diamond edges become more
smeared out, but in contrast to the behavior in the $18~\um^2$ device, periodic
oscillations as a function of dc bias are not seen.

\begin{figure}[h]
\center \label{figure4}
\includegraphics[width=2.9in]{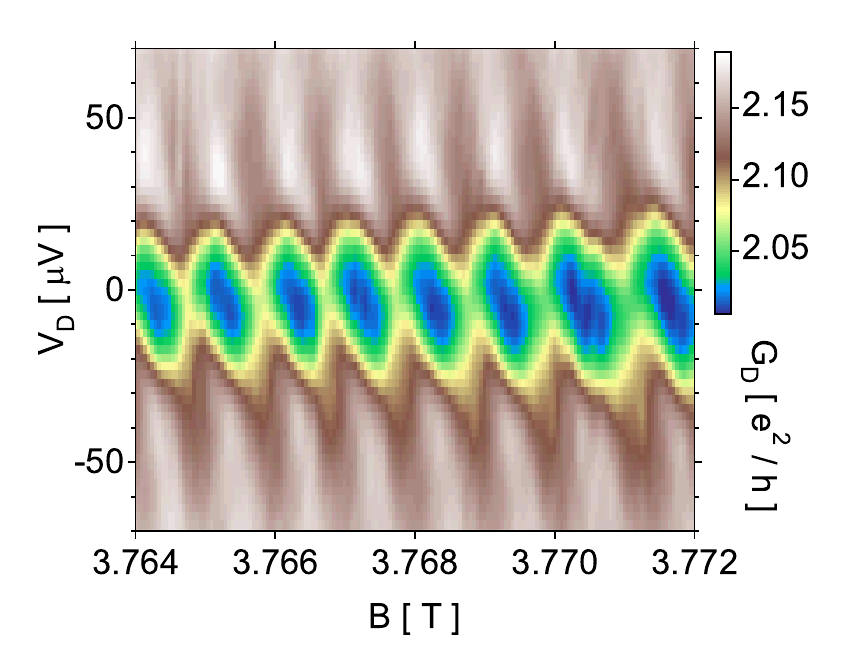}
\caption{\footnotesize{$\gd$ as a function of $B$ and $\vd$ in the $2~\um^2$ device.}}
\end{figure}

In conclusion, quantum Hall FPI's large enough that Coulomb charging is negligible
are found to display both AB and \FP conductance oscillations. The
combination of these two effects yields a checkerboard-like pattern of
oscillations from which the edge-state velocity and dephasing rate can be
extracted, and both are found to be consistent with theoretical calculations.
Although this pattern resembles that seen in Mach-Zehnder interferometers, the
dependence of its characteristics on magnetic field is evidently quite different
from what has been observed in those devices~\cite{neder06,litvin08}, providing
experimental evidence that the underlying mechanisms for oscillations with bias in
the two types of devices may be quite different.

We are grateful to R.\ Heeres for technical assistance and to B.\ I.\ Halperin,
M.\ A.\ Kastner, C.\ de~C.\ Chamon, A.\ Stern, I. Neder, R.\ Gerhardts, J.\ B.\ Miller and
I.\ P.\ Radu for enlightening discussions. This research has been funded in part
by Microsoft Corporation Project Q, IBM, NSF (DMR-0501796), Harvard University,
and the Heisenberg program of DFG. Device fabrication at Harvard Center for Nanoscale Systems.

\small
\bibliographystyle{dougprl}
\bibliography{FQHE}
\end{document}